\documentclass[10pt,twocolumn,english,pre]{revtex4-1}
\usepackage[T1]{fontenc}
\usepackage[latin9]{inputenc}
\usepackage{geometry}
\usepackage{color}
\usepackage{amsmath}
\usepackage{amssymb}
\usepackage{graphicx}

\makeatletter
\usepackage{algpseudocode}

\makeatother

\usepackage{babel}
\usepackage{listings}
\begin{document}
\global\long\def\ee{\mathrm{e}}%
\global\long\def\ii{\mathrm{i}}%
\global\long\def\iv{\mathrm{i}\nu}%
\global\long\def\ivp{\mathrm{i}\nu^{\prime}}%
\global\long\def\iw{\mathrm{i}\omega}%
\global\long\def\iW{\mathrm{i}\Omega}%
\global\long\def\eye{\boldsymbol{1}}%
\global\long\def\bias{\operatorname{bias}}%
\global\long\def\EE{\operatorname{E}}%
\global\long\def\Var{\operatorname{Var}}%
\global\long\def\thVar{\mathrm{\widehat{\Var}}}%
\global\long\def\Cov{\operatorname{Cov}}%
\global\long\def\gap{\operatorname{gap}}%
\global\long\def\tri{\operatorname{tri}}%
\global\long\def\tauint#1{\tau_{\mathrm{int},#1}}%
\allowdisplaybreaks
\title{Efficient estimation of autocorrelation spectra}
\author{Markus Wallerberger}
\affiliation{Department of Physics, University of Michigan, Ann Arbor, MI 48109,
USA}
\begin{abstract}
The performance of Markov chain Monte Carlo calculations is determined
by both ensemble variance of the Monte Carlo estimator and autocorrelation
of the Markov process. In order to study autocorrelation, binning
analysis is commonly used, where the autocorrelation is estimated
from results grouped into bins of logarithmically increasing sizes.

In this paper, we show that binning analysis comes with a bias that
can be eliminated by combining bin sizes. We then show binning analysis
can be performed on-the-fly with linear overhead in time and logarithmic
overhead in memory with respect to the sample size. We then show that
binning analysis contains information not only about the integrated
effect of autocorrelation, but can be used to estimate the spectrum
of autocorrelation lengths, yielding the height of phase space barriers
in the system. Finally, we revisit the Ising model and apply the proposed
method to recover its autocorrelation spectra.
\end{abstract}
\maketitle

\section{Introduction}

The theoretical treatment of problems at the forefront of physics
research often involves integration of a highly non-trivial function
over a ma\-ny-di\-men\-sio\-nal space. Prominent examples include
condensed matter with strong electronic correlation~\citep{loewdin-acp07},
interacting nuclear matter, lattice quantum field theories, as well
as systems with geometrical frustration. For these problems, brute
force attempts at a solution are usually met with an ``exponential
wall'', i.e., a cost in resources that rises exponentially with the
complexity of the system under study.

Markov chain Monte Carlo (MCMC) techniques evaluate the integral stochastically
by performing a random walk in the high-dimensional space~\citep{LandauBinder05,krauth-statmech06}.
They are ideally suited for high-dimensional integrals, and have led
to many breakthroughs in the aforementioned fields. Nevertheless,
major generic challenges such as infinite variances~\citep{shi-pre16}
and the fermionic sign problem~\citep{loh-prb-1990} remain, and
this has led the community to create more and more elaborate Markov
chain Monte Carlo algorithms~\citep{gubernatis-latticemc16}. In
order to do so, it is important to be able to analyze the statistical
properties of the algorithms we create. Otherwise, we can only rely
on our physical intuition and cumbersome trial-and-error to improve
our method.

Due to the remarkable simplicity of the central limit theorem, the
performance of a plain Monte Carlo integration is only affected by
the variance of the integrand, which we can trivially estimate. We
can use this in an attempt to construct ``improved estimators''
or to rebalance the deterministic and stochastic part of the algorithm~\citep{liu-mc01}.

However, as soon as we sample along a Markov chain, the performance
is now a combination of the variance of the integrand and the total
effect of autocorrelation along the Markov chain~\citep{sokal-bosonic92}.
In order to analyze and improve our algorithm, we must disentangle
the properties of the Markov chain, which relates to the efficiency
of the moves we chose, from the properties of the space under study.

Two techniques are in wide use to estimate autocorrelation: (i) direct
measurement of the autocorrelation function~\citep{madras-sokal-jsp88,janke-stat02}
and (ii) binning or blocking analysis\citep{flyvbjerg-jcp91}. Measurement
of the autocorrelation function is straightforward, and allows one
to analyze the structure of the Markov chain in more detail. However,
it is also expensive in terms of time and memory, and in order for
estimators to have finite variance, it usually requires an \emph{ad
hoc }regularization of the sum, potentially introducing bias. Binning
analysis, on the other hand, can be performed considerably faster
and with less memory. However, it only provides an estimate for the
integrated effect of autocorrelation.

In this paper, we first show that the logarithmic binning analysis
(Sec.~\ref{sec:binningrev}), while asymptotically correct, has a
practical bias. We show a simple trick to correct for that bias (Sec.~\ref{sec:biascorr}).
We then show how to perform the binning analysis with only a constant
overhead in time per move (Sec.~\ref{sec:fast}). We then show that
the logarithmic binning analysis can not only be used to gauge the
total effect of autocorrelation, but can also be used to estimate
the height and strength of phase space barriers present in the system
(Sec.~\ref{sec:spectrum}) and illustrate the method on the Ising
model (Sec.~\ref{sec:ising}).

\section{Review: Markov chain spectra\label{sec:review}}

In order to move forward with the algorithm, it is useful to review
a couple of key points on Markov chains and Monte Carlo integration
along Markov chains~\citep{grimmett-markov12}. For simplicity, we
will focus on finite reversible Markov chains, but the results are
equally applicable to infinite and non-reversible chains. Unless explicitly
noted, we only consider stationary Markov chains (see later).

A Markov chain is characterized by a number $S$ of states and an
$S\times S$ transition matrix $P$ encoding the probabilities $P_{xy}\in[0,1]$
to go from state $y$ to state $x$. We require the Markov chain to
be ergodic (irreducible), which roughly means that each state can
be reached from any other state in a finite number of steps. Each
ergodic Markov chain tends to a stationary distribution $\pi$. We
would like to choose $P$ in such a way that the stationary distribution
matches the distribution we would like to sample. For this, it is
sufficient to require the Markov chain to be in detailed balance (reversibility)
with respect to $\pi$; formally: $P_{xy}\pi_{y}=P_{yx}\pi_{x}$.

Using the Perron--Frobenius theorem, one can show that a Markov chain
satisfying above criteria has the following spectral decomposition~\citep{grimmett-markov12}:
\begin{equation}
P_{xy}=\pi_{x}+\sum_{i=1}^{S-1}\alpha_{i}\sqrt{\pi_{x}}b_{i,x}b_{i,y}\frac{1}{\sqrt{\pi_{y}}},\label{eq:spectrum}
\end{equation}
where $-1<\alpha_{i}<1$ are eigenvalues and $\{\sqrt{\pi},b_{1},\allowbreak b_{2},\ldots\}$
are a set of vectors forming an orthonormal basis. We see that indeed
$\pi$ is a stationary distribution of $P$, as $\pi$ is a right
eigenvector of $P$ with eigenvalue 1: $P\pi=\pi$. However, there
are additional modes in Eq.~(\ref{eq:spectrum}) that decay on a
characteristic scale $\tau_{i}$ in Markov time, where
\begin{equation}
|\alpha_{i}|=\exp(-1/\tau_{i}).\label{eq:taui}
\end{equation}
The $\alpha_{\exp}=\max_{i}|\alpha_{i}|$ determines the so-called
spectral gap of the Markov chain, defined as $(1-\alpha_{\mathrm{exp}})$.
The corresponding $\tau_{\exp}$ gives a worst-case estimate for how
many steps it takes a Markov chain to reach a stationary distribution.
It thus provides an important lower bound for the number of initial
steps that should be discarded in a simulation, and is sometimes called
exponential autocorrelation time~\citep{sokal-bosonic92,sokal-lecture96}.

Now let us consider random variable $A$ with expectation value $\langle A\rangle=\sum_{y}A_{y}\pi_{y}$.
We would like to estimate $A$ using a suitable Markov chain. To do
so, we construct the Markov chain Monte Carlo estimator $\hat{A}$
by sampling $N$ steps $A^{(0)},\ldots,A^{(N-1)}$ along the Markov
chain, i.e.,
\begin{equation}
\hat{A}:=\frac{1}{N}\sum_{n=0}^{N-1}A^{(n)}\quad\mathrm{with}\quad A_{y}^{(n)}:=\sum_{x}A_{x}P_{xy}^{n}.\label{eq:mcest}
\end{equation}
If we have sufficiently thermalized the run, we sample according to
the stationary distribution $\pi$, and the estimator $\hat{A}$ provides
an unbiased estimate for the expectation value of $A$, i.e., $\langle\hat{A}\rangle=\langle A\rangle$.
For the variance, however, one finds~\citep{sokal-lecture96}:
\begin{equation}
\Var\hat{A}=\frac{1}{N}\sum_{p=-N}^{N}\left(1-\frac{|p|}{N}\right)c_{A}(p)\Var A,\label{eq:varahat}
\end{equation}
where $c_{A}(p)$ is the autocorrelation function of $A$ and distance
or ``lag'' $p$ in Markov time, defined as:
\begin{align}
c_{A}(p) & :=\frac{\langle A^{(n)}A^{(n+p)}\rangle-\langle A\rangle^{2}}{\langle A^{2}\rangle-\langle A\rangle^{2}}=\sum_{i=1}^{S-1}\alpha_{i}^{|p|}\frac{\langle Ab_{i}^{\prime}\rangle^{2}}{\Var A},\label{eq:Cap}
\end{align}
where $b_{i,x}^{\prime}=b_{i,x}/\sqrt{\pi_{x}}$. The autocorrelation
function starts at 1 for $p=0$, indicating that a measurement is
always perfectly correlated with itself, and drops off exponentially
in both directions. The decay again depends on the characteristic
decay times (\ref{eq:taui}). However, the coefficient $\langle Ab_{i}\rangle^{2}$
depends on the influence of the corresponding ``mode'' of the Markov
chain on the quantity under study.

Taking the limit $N\to\infty$ in Eq.~(\ref{eq:varahat}), we find
the familiar expression~\citep{sokal-bosonic92,liu-mc01,berg-mcmc04}:
\begin{equation}
\Var\hat{A}\to\frac{\tauint A}{N}\Var A,\label{eq:largenumbers}
\end{equation}
where we have defined the integrated autocorrelation time\emph{ }$\tauint A$
for $A$, given by:\footnote{Another common convention is to define $\tauint A$ as the sum of
$p=1,\ldots,\infty$ of the autocorrelation function. In this case,
the right-hand side of Eq.~(\ref{eq:tauint}) is 2$\tauint A$+1.}
\begin{equation}
\tauint A:=\sum_{p=-\infty}^{\infty}c_{A}(p)=\sum_{i=1}^{S-1}\frac{1+\alpha_{i}}{1-\alpha_{i}}\cdot\frac{\langle Ab_{i}^{\prime}\rangle^{2}}{\Var A}.\label{eq:tauint}
\end{equation}
The integrated autocorrelation time is the total effect of autocorrelation
on the variances. Comparing Eq.~(\ref{eq:largenumbers}) with the
central limit theorem for non-autocorrelated samples, we can say autocorrelation
reduces the number of independent samples from $N$ to $N^{\prime}=N/\tauint A$.

Eq.~(\ref{eq:largenumbers}) makes it clear why the estimating the
autocorrelation time is important when designing a new Monte Carlo
algorithm: the variance of the estimate is a combination of the intrinsic
efficiency of the Monte Carlo procedure, which only depends on the
quantity and the space under study, and the efficiency of the Markov
chain in exploring said space. These two things need to be disentangled
in order to gauge whether we need to improve the quality of our proposed
moves or work on improving the estimator itself.

\section{Logarithmic Binning Analysis\label{sec:binning}}

\subsection{Review: Autocorrelation estimates\label{sec:binningrev}}

One strategy to estimate the integrated autocorrelation time is to
turn Eqs. (\ref{eq:Cap}) and (\ref{eq:tauint}) into an estimator.
One usually first constructs an estimator for the autocovariance function~\citep{sokal-lecture96}:
\begin{equation}
\hat{C}_{A}(p)=\frac{1}{N-|p|-1}\!\sum_{n=0}^{N-|p|-1}(A^{(n)}-\hat{A})(A^{(n+|p|)}-\hat{A}).\label{eq:Cest}
\end{equation}
By normalizing the autocovariance function we obtain an estimator
for the autocorrelation function:
\begin{equation}
\hat{c}_{A}(p):=\frac{\hat{C}_{A}(p)}{\hat{C}_{A}(0)},\label{eq:cest}
\end{equation}
which we can sum up to obtain an estimator for the integrated autocorrelation
time:
\begin{equation}
\hat{\tau}_{\mathrm{int},A}=\sum_{p=-\infty}^{\infty}\omega(p)\hat{c}_{A}(p),\label{eq:tauintsum}
\end{equation}
where we have introduced a window function $\omega(p)$ that should
in principle be taken as $\omega(p)=1$.

In practice, there are two problems with this approach: first, the
noise-to-signal ratio in the autocorrelation estimator blows up as
\begin{equation}
\frac{\Var\hat{c}_{A}(p)}{c_{A}(p)}=\mathcal{O}\left(\frac{\exp(|p|/\tau_{\mathrm{max}})}{N-|p|}\right),\label{eq:signaltonoise}
\end{equation}
as the autocorrelation function drops exponentially with some time
scale $\tau_{\mathrm{max}}$, while only $(N-|p|)$ data points are
available in Eq.~(\ref{eq:Cest}). Plugging this into Eq.~(\ref{eq:tauintsum}),
the variance of the sum diverges. To remedy that one has to choose
a hard cutoff $W$ in the window function $\omega(p)$, truncating
the sum for $|p|>W$, introducing bias~\citep{madras-sokal-jsp88,janke-stat02}.
Secondly, the procedure is computationally expensive, requiring $\mathcal{O}(WN)$
of time and $\mathcal{O}(W)$ of memory in the case of a cutoff chosen
\emph{a priori}, or $\mathcal{O}(N\log N)$ time and $\mathcal{O}(N)$
of memory if the cutoff is determined \emph{a posteriori}.

The other strategy is to construct a binning estimator $\hat{A}^{(M)}$:
instead of sampling over $N$ individual measurements, we group them
into $B$ non-over\-lapping bins $A^{(M,0)},\allowbreak A^{(M,1)},\ldots,A^{(M,B-1)}$.
Each bin is the mean of $M=N/B$ consecutive measurements, so the
zeroth bin $A^{(M,0)}$ contains the mean of the inital $M$ samples,
$A^{(0)}\ldots A^{(M-1)}$; the first bin $A^{(M,1)}$ contains the
mean of the next $M$ samples; and so on: 
\begin{equation}
\hat{A}^{(M)}:=\frac{1}{B}\sum_{b=0}^{B-1}A^{(M,b)};\;A^{(M,b)}:=\frac{1}{M}\sum_{m=0}^{M-1}A^{(bM+m)}.\label{eq:AM}
\end{equation}
Comparing Eqs. (\ref{eq:AM}) and Eq.~(\ref{eq:mcest}), we find
that each bin can be understood as a small sampling procedure with
the sample size $N$ replaced by the bin size $M$. Thus, for large
enough $M$, the variance of the individual bin means $A^{(M,b)}$
is given by Eq.~(\ref{eq:largenumbers}), which we can estimate from
the data since we have $B$ bins:
\begin{equation}
\widehat{\Var}A^{(M)}=\frac{1}{B-1}\sum_{b=0}^{B-1}\big[A^{(M,b)}-\hat{A}^{(M)}\big]^{2}.\label{eq:varest}
\end{equation}
(We emphasize the difference in notation: $\Var\hat{A}$ is the variance
of the mean estimator, or in other words the squared standard error
of the mean, while {\small{}$\widehat{\Var}\,$}$A$ is an estimator
of the variance.)

Thus, we can turn Eq. (\ref{eq:largenumbers}) into a binned estimator
for the integrated autocorrelation time~\citep{berg-mcmc04}:
\begin{equation}
\hat{\tau}_{\mathrm{int},A}^{(M)}:=\frac{M\cdot\widehat{\Var}A^{(M)}}{\widehat{\Var}A^{(1)}}.\label{eq:tauest}
\end{equation}
In the limit $M\to\infty$, Eq.~(\ref{eq:largenumbers}) trivially
ensures that the bias, defined as:
\begin{equation}
\bias\hat{\tau}_{\mathrm{int},A}:=\langle\hat{\tau}_{\mathrm{int},A}\rangle-\tauint A\label{eq:bias}
\end{equation}
vanishes. For a finite and fixed number of simulation steps $N$,
however, one has to make a trade-off between bias, which improves
with $\mathcal{O}(1/M)$, and the statistical error of the integrated
autocorrelation time estimator (\ref{eq:tauest}), which increases
as $\mathcal{O}(\sqrt{M})$.

\begin{figure}
\includegraphics[width=1\columnwidth]{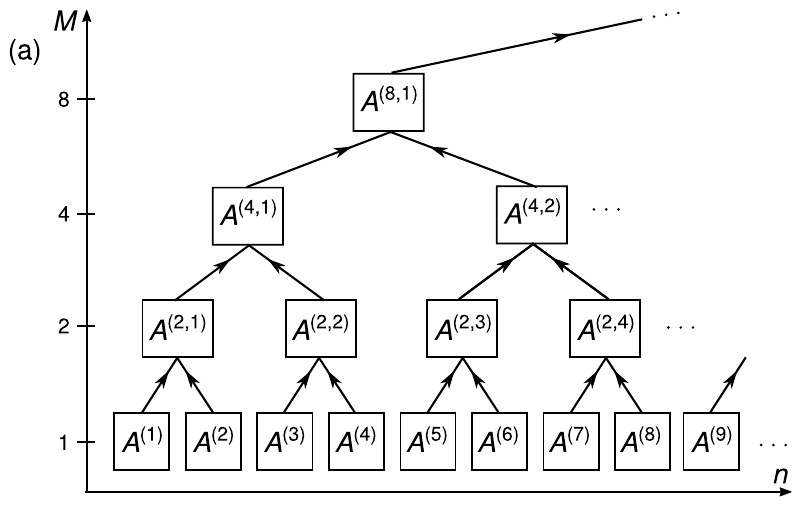}

\includegraphics[width=1\columnwidth]{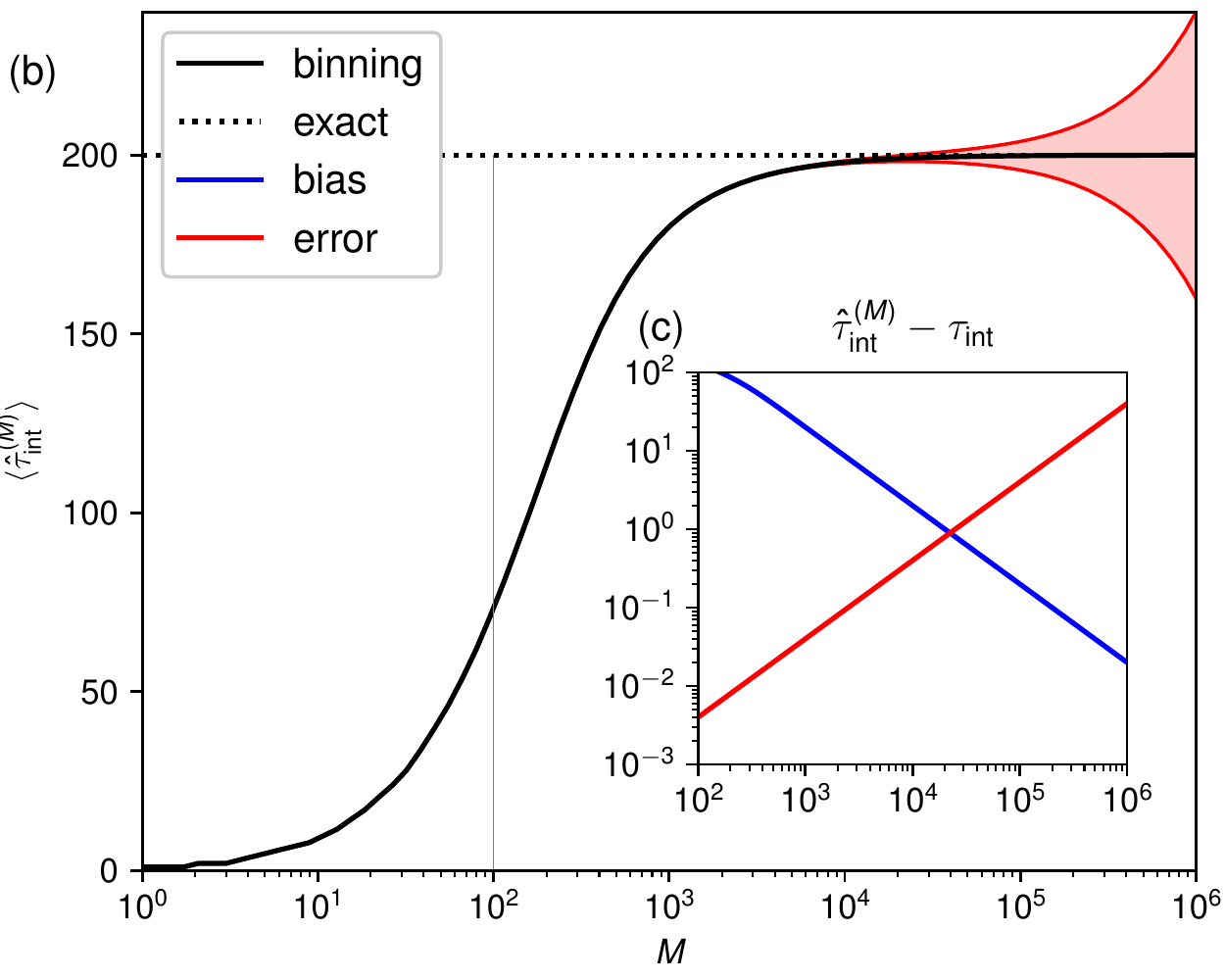}

\caption{Logarithmic binning analysis. (a) Illustration of the procedure for
combining two lower-level bin means to a higher-level bin via Eq.~(\ref{eq:bincombine});
(b) Estimates for the autocorrelation time for a single $\tau=100$
dependent on $M$ using Eq.~(\ref{eq:tauest}) with an estimate for
the statistical error; (c) error in the estimate due to bias (blue
line) and statistical uncertainty (red line).}
\label{fig:binning}
\end{figure}

Because a single, \emph{ad hoc }choice for $M$ makes it difficult
to gauge this tradeoff, one usually performs what is known as \emph{blocking}
or (\emph{logarithmic}) \emph{binning analysis}~\citep{flyvbjerg-jcp91}.
There, one computes {\small{}$\widehat{\Var}\,$}$A^{(M)}$ for a
set of logarithmically spaced $M=1,2,4,8,\ldots,N$. This can be done
efficiently by combining two lower-level bins into higher-level bins
according to:
\begin{equation}
A^{(2M,b)}=\tfrac{1}{2}(A^{(M,2b)}+A^{(M,2b+1)}),\label{eq:bincombine}
\end{equation}
which, for a precomputed set of measurements, can be performed in
$\mathcal{O}(N)$ time and $\mathcal{O}(\log N)$ of additional memory
(see Fig.~\ref{fig:binning}a).

One then plots the autocorrelation estimate (\ref{eq:tauest}) over
$\log M$ and assesses the trade-off, usually by visual inspection.
As an example, let us take an autocorrelation function with a single
time scale $\tau=100$. The mean of the autocorrelation estimator
(\ref{eq:tauint}) for different values of $M$ is shown as solid
black line in Fig.~\ref{fig:binning}b: we see that for $M>\tau$,
$\langle\tau_{\mathrm{int}}\rangle$ approaches the true value of
$\tau$, represented as dotted black line. (Note that $\tauint A\approx200$
since both the effect of autocorrelation with negative and positives
offsets are included in our definition.) At the same time, the statistical
error starts to grow, signified by the red interval~\citep{gubernatis-latticemc16}.

\subsection{Bias correction\label{sec:biascorr}}

Fig.~\ref{fig:binning}c makes the trade-off between systematic bias
(blue), computed from Eq.~(\ref{eq:bias}), and statistical error
(red), estimated from the asymptotic expression, in the logarithmic
binning analysis explicit. We see that the autocorrelation estimate
converges relatively slowly, $\mathcal{O}(1/M)$, to the true result.
Since the statistical error rises relatively quickly, $\mathcal{O}(M)$,
as well, it can be difficult to find a good tradeoff between the two
errors.

\begin{figure}
\includegraphics[width=1\columnwidth]{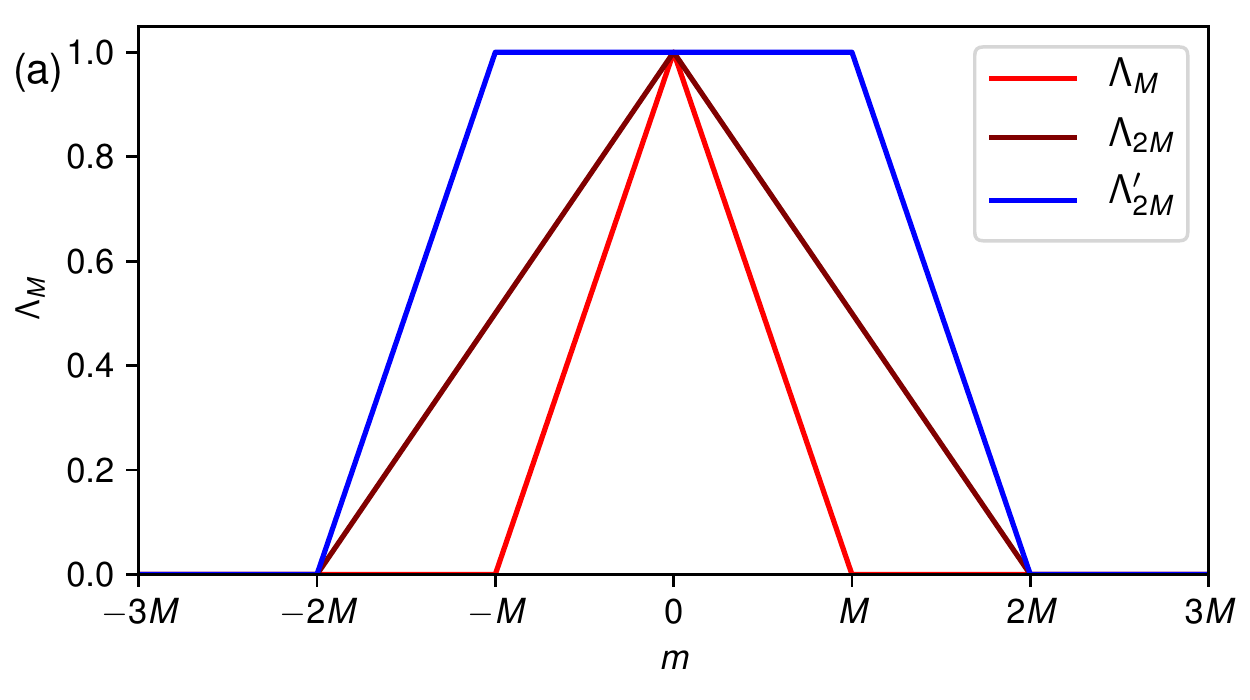}

\includegraphics[width=1\columnwidth]{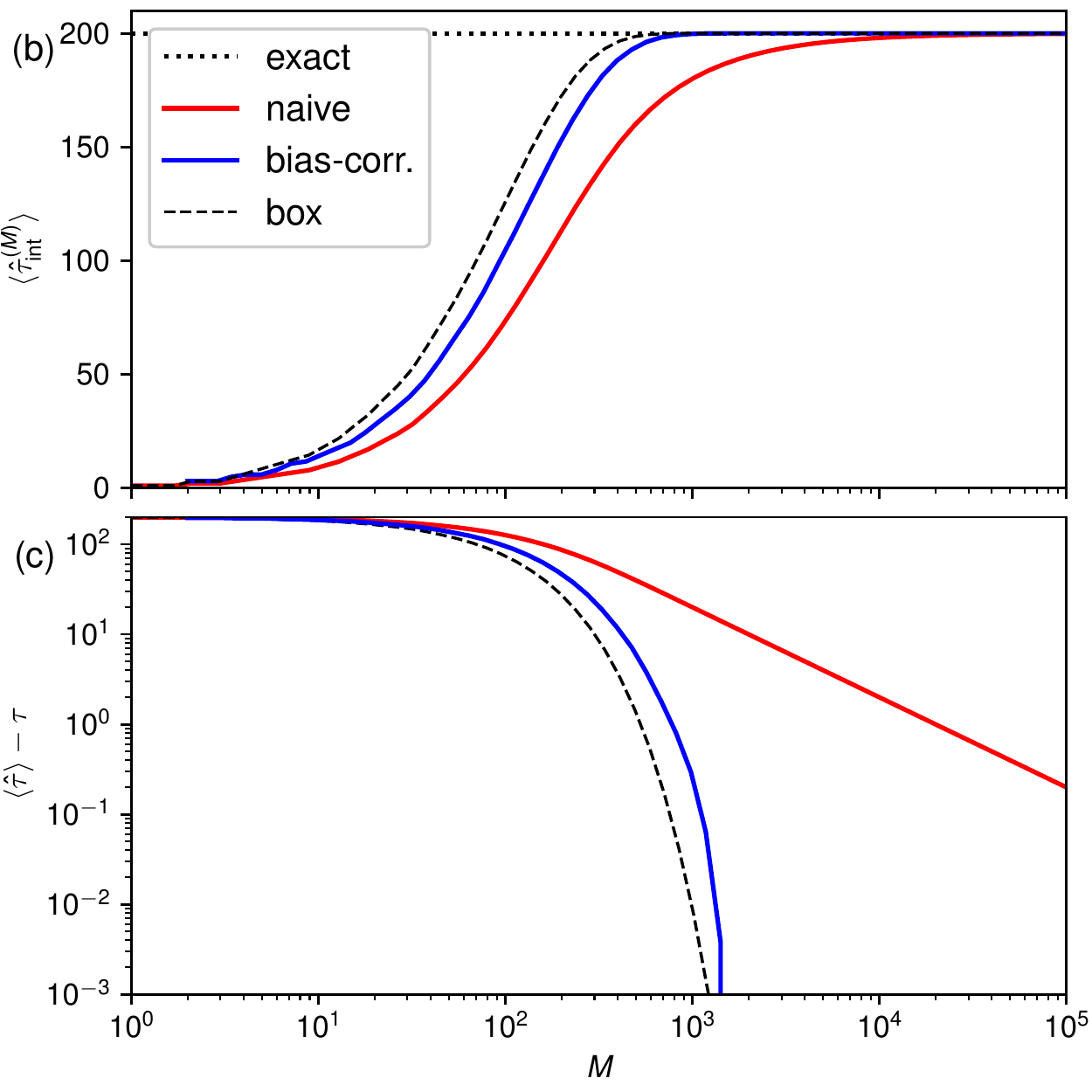}

\caption{(a) value of the triangular window functions $\Lambda_{M}$ (red curves)
as well as the trapezoid window function $\Lambda_{2M}^{\prime}$
(blue curve); (b) Naive (red curve) as well as bias-corrected (blue
curve) estimates for the autocorrelation time for a single $\tau=100$
for different values of $M$; (c) bias (\ref{eq:bias}) as a function
of $M$. For comparison, the full summation up of the autocorrelation
function up to $|m|=M$ is shown as black dashed line.}

\label{fig:tauimpr}
\end{figure}

To understand the bias, let us consider the expectation value for
the binned variance estimator (\ref{eq:varest}):
\begin{equation}
\begin{split}\langle\widehat{\Var}A^{(M)}\rangle & =\frac{1}{B}\sum_{b}\frac{1}{M^{2}}\sum_{m,n}\langle A^{(bM+m)}A^{(bM+n)}\rangle\\
 & -\frac{1}{B^{2}}\sum_{b,b'}\frac{1}{M^{2}}\sum_{m,n}\langle A^{(bM+m)}\rangle\langle A^{(b'M+n)}\rangle.
\end{split}
\label{eq:varexpr}
\end{equation}
Using Eq.~(\ref{eq:mcest}) and employing the stationarity condition
(we have ``thermalized'' the run), one can write Eq.~(\ref{eq:varexpr})
in terms of the autocorrelation function (\ref{eq:Cap}) as follows:
\begin{align}
\langle\widehat{\Var}A^{(M)}\rangle & =\frac{\Var A}{M^{2}}\sum_{m,n=0}^{M-1}c_{A}(m-n)\nonumber \\
 & =\frac{\Var A}{M}\sum_{m=-\infty}^{\infty}\Lambda_{M}(m)c_{A}(m),\label{eq:chiMcA}
\end{align}
where $\Lambda_{M}(m)$ is the scaled triangular window function of
width $2M$:
\begin{equation}
\Lambda_{M}(m):=\begin{cases}
1-\frac{|m|}{M} & |m|\le M\\
0 & \mathrm{else}
\end{cases}.\label{eq:chiM}
\end{equation}
The values of the window for $M$ and $2M$ are plotted in Fig.~\ref{fig:tauimpr}a
(red curves). The convolution of the autocorrelation function with
the window expresses the fact that we discard correlations that cross
bin boundaries. (Those correlations could be encoded by constructing
a covariance matrix across bins or bin sizes.)

Examining Eq.~(\ref{eq:chiM}), we see that the autocorrelation time
estimator Eq.~(\ref{eq:tauest}) must have a significant bias, as
we already cut away some of the short-length autocorrelation, e.\,g.,
the correlation of the last measurement of the $b$'th bin with the
first measurement of the $(b+1)$'st bin.

However, we can use the fact that we have logarithmically spaced bins
to our advantage and construct a trapezoid window function by linear
combination:
\begin{equation}
\Lambda_{2M}^{\prime}(m):=2\Lambda_{2M}(m)-\Lambda_{M}(m).\label{eq:lambdap}
\end{equation}
Such a new window function is plotted as blue curve in Fig.~\ref{fig:tauimpr}a:
we see that we take autocorrelations up to length $M$ fully into
account, and include diminishing contributions from lengths $M$ to
$2M$. The corresponding \emph{bias-corrected autocorrelation estimator}
is given by:
\begin{equation}
\hat{\tau}_{\mathrm{int},A}^{\prime(2M)}:=\frac{4M\cdot\widehat{\Var}A^{(2M)}-M\cdot\widehat{\Var}A^{(M)}}{\widehat{\Var}A^{(1)}}.\label{eq:tauprime}
\end{equation}

Fig.~\ref{fig:tauimpr}b compares the naive estimator (red curve)
using Eq.~(\ref{eq:tauest}) with the bias-corrected estimator (blue
curve) using Eq.~(\ref{eq:tauprime}), again for a single autocorrelation
mode with $\tau=100$. For comparison, the explicit summation (\ref{eq:tauint})
of the autocorrelation function with a hard cutoff $|m|<M$ is shown
as black dashed curve. We see that the bias-corrected estimator approaches
the exact value of $\tau_{\mathrm{int}}$ nearly as fast as the explicit
summation.

Examining the bias (Fig.~\ref{fig:tauimpr}c), we find that while
the bias of the naive estimator drops as $\mathcal{O}(\tau_{\mathrm{max}}/M)$,
the bias of the corrected estimator drops as $\mathcal{O}(\exp(-M/\tau_{\mathrm{max}}))$,
as we include all short-range autocorrelation contributions explicitly.
This results in a stable and pronounced ``plateau'' in the improved
estimator, the height of which yields the integrated autocorrelation
time.

The statistical error of both naive and improved estimator scale as
$\mathcal{O}(\sqrt{M/N})$, but the reduction of bias comes at the
cost of an at most two-fold increase in the statistical error. This
is a beneficial tradeoff: suppose we want to choose $M(N)$ such that
bias and stastical error have the same asymptotic behaviour with $N\to\infty$.
For the naive estimator, we should choose $M\propto N^{1/3}$, which
yields a total error of $\mathcal{O}(N^{-1/3})$~\citep{frezzotti-qcd}.
For the improved estimator, we can choose $M=\tau_{\max}+c\log(N)$,
where $c>0$ is a parameter, which improves the total error to $\mathcal{O}(N^{-1/2})$.
In the case where $\tau_{\max}$ is unknown, the choice $M\propto N^{\gamma}$
with $\gamma\in(0,1/3)$ still yields better asymptotic behavior than
in the naive case.

\subsection{Example: vector autoregression\label{sec:var1-tauint}}

To illustrate and test the effect of bias correction, we use a first-order
vector autoregression model (VAR(1) model). One of the simplest autocorrelated
Markov processes, the VAR(1) model can be written as~\citep{enders-ts04}:
\begin{equation}
X^{(t)}=\phi_{0}+\phi_{1}X^{(t-1)}+E^{(t)},\label{eq:var1}
\end{equation}
where $X^{(t)}$ is a random $n$-vector for the $t$-th step of the
process, $\phi_{0}$ is a constant vector, $\phi_{1}$ is an $n\times n$
matrix, and $E^{(t)}$ is a set of random vectors identically and
independently distributed according to $\mathcal{N}(0,\Sigma_{E})$,
introducing white noise into the system~\footnote{We note that the VAR(1) model has an infinite state space and thus
the Markov chain review of Sec.~\ref{sec:review} is not directly
applicable. However, the setup is easily generalized to infinite state
spaces for which all the conclusions in this paper remain valid.}.

For $||\phi_{1}||<1$, the process will tend to a stationary case.
The autocorrelation function is then simply given by~\citep{enders-ts04}:
\begin{equation}
c_{X}(m)=\phi_{1}^{|m|}\label{eq:cXvar1}
\end{equation}
and the covariance matrix of $X$ is the solution to the following
Lyapunov (matrix) equation:
\begin{equation}
\Cov X-\phi_{1}(\Cov X)\phi_{1}^{\mathrm{T}}=\Sigma_{E}.\label{eq:sylv}
\end{equation}

VAR(1) models are most commonly used as a fitting method for autocorrelated
observations~(cf.~Sec.~\ref{sec:spectrum-review}). We are going
to run it in the opposite direction and use Eq.~(\ref{eq:var1})
as prescription for generating an autocorrelated dataset. We used
two components ($n=2$) and generated the propagation matrix $\phi_{1}$
from two eigenvalues $\alpha=(0.9,0.985)$ and rotated the eigenbasis
by $60$ degrees to introduce coupling between the components. We
then generated $10$ runs of $N=2^{24}$ configurations each, and
measured the first component $X_{1}^{(t)}=:Y$ for each configuration.

The integrated autocorrelation is then given analytically from Eq.~(\ref{eq:cXvar1}):
\begin{equation}
\tauint Y=\frac{[(\eye+\phi_{1})(\eye-\phi_{1})^{-1}\Cov X]_{11}}{[\Cov X]_{11}},\label{eq:tauinty}
\end{equation}
where $[...]_{11}$ denotes the $(1,1)$-component of the respective
matrix. (The covariance appears in Eq.~(\ref{eq:tauinty}) because
when we truncate $X$ to form the observable $Y$, we silently discard
the cross-correlations in the normalization of the autocovariance
function.)

\begin{figure}
\includegraphics[width=1\columnwidth]{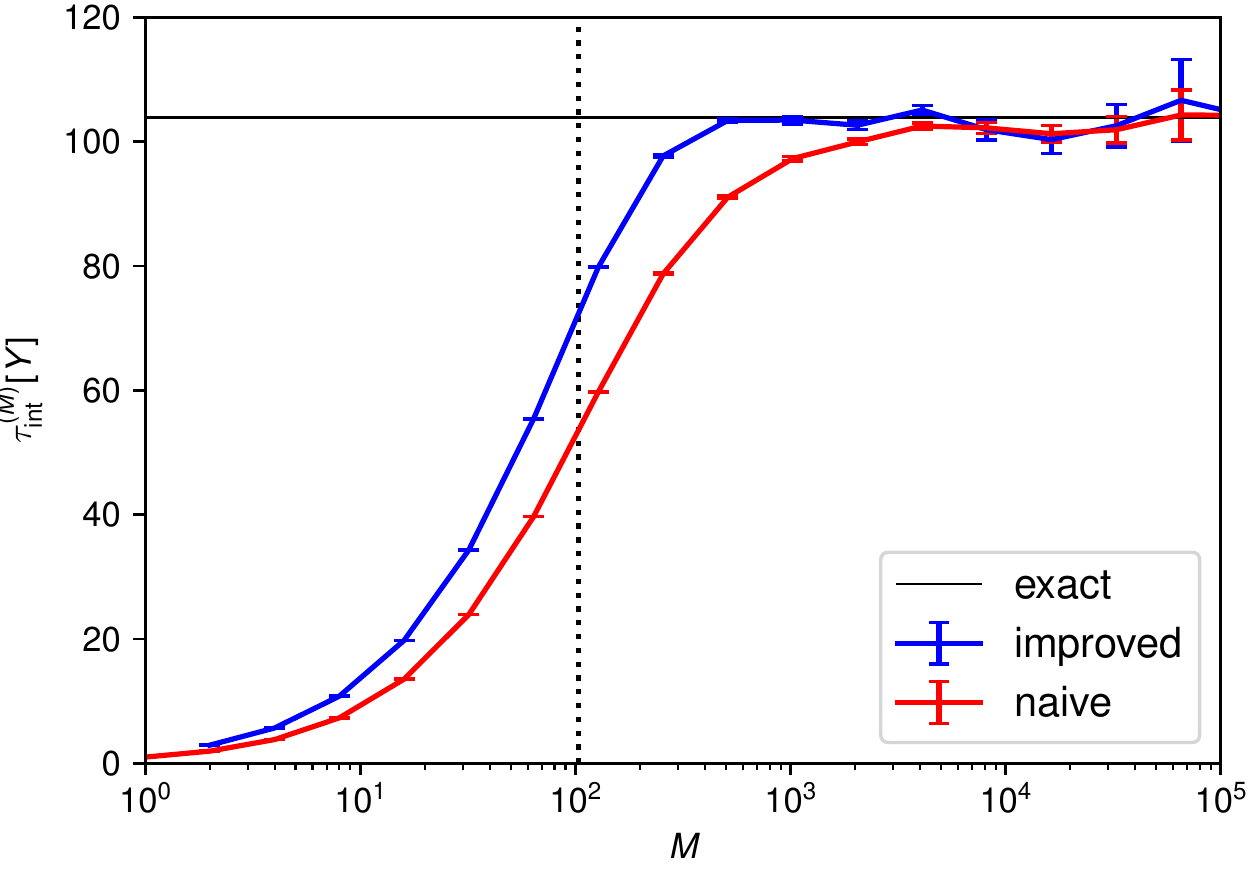}

\caption{Logarithmic binning analysis for the first component $Y=X_{1}$ a
two-component VAR(1) model. Naive (red line) and bias-corrected (blue
line), logarithmic binning estimator for $\protect\tauint Y$, where
mean and standard error are computed from 10 runs of $N=2^{24}$ measurements
each. The analytic result from Eq.~(\ref{eq:tauinty}) is shown as
black solid line.}
\label{fig:var1-tauint}
\end{figure}

Fig.~\ref{fig:var1-tauint} compares results from the naive estimator
(\ref{eq:tauest}) with ones from the bias-corrected estimator (\ref{eq:tauprime})
for the autocorrelation time of the VAR(1) model outlined above. The
exact value for $\tauint Y\approx103.91$ from Eq.~(\ref{eq:tauinty})
is shown as black solid line as well as black dashed line. As expected
from Sec.~\ref{sec:biascorr}, the naive estimator (red curve) significant
bias that persists for several orders of magnitude of $M$ even after
$M$ exceeds the integrated autocorrelation time. For large $M$,
where the bias decays, its effect intermingles with the statistcal
uncertainty, which renders the extraction of the autocorrelation time
unreliable.

The improved estimator (blue curve), on the other hand, converges
exponentially to a stable plateau as soon as $M$ exceeds $\tauint Y$.
By extracting the height of the plateau one obtains a reliable estimate
for $\tauint Y$.

As discussed in Sec.~\ref{sec:biascorr}, we see that the improved
estimator has statistical error bars that are about twice as large
as the naive estimator, but the exponential drop of the bias allows
us to choose a lower $M$, which yields a better estimate of the plateau
height and thus $\tauint Y$.

\subsection{Online estimation in linear time\label{sec:fast}}

Given a Monte Carlo run with $N$ measurements, we can perform the
logarithmic binning analysis in $\mathcal{O}(N)$ time: computing
the variance over $B$ bins and combining bins both take $\mathcal{O}(B)$
time, and the number of bins shrinks by a factor of $2$ each time
we double the bin size. We find:
\begin{equation}
\mathcal{O}\left[N+\frac{N}{2}+\frac{N}{4}+\ldots\right]=\mathcal{O}(2N).\label{eq:complex}
\end{equation}
For storing the variances for the different $M=1,2,4,\ldots$, we
need $\mathcal{O}(\log N)$ of memory. However, we also need to store
the measurements in order to perform the binning analysis on them.
This takes an additional $\mathcal{O}(N)$ of memory, which grows
prohibitively quickly for large runs.

We would like to eliminate this memory requirement, and are thus looking
for an algorithm that can perform the analysis \emph{online}, i.e.,
processing measurements one at a time without the need to store the
whole data series.

One can solve this problem by first realizing that the variance can
be estimated online~\citep{knuth-aocp2,chan-stat83}: in the simplest
case, we initialize three accumulators $s,S_{1},S_{2}\gets0$. Each
data point $x_{i}$ is first added to the bin accumulator $s\gets s+x_{i}$.
Once we collected $M$ measurements, we update the accumulator according
to $S_{p}\gets S_{p}+s^{p}$ and reset the bin $s\gets0$. At the
end, we combine the results to the variance~\footnote{We note here that the separate accumulation of $S_{1}$ and $S_{2}$
may cause stability problems; more advanced accumulation techniques
can be used instead at little additional overhead.\citep{chan-stat83}}. We can therefore do a binning analysis by setting up a triplet of
accumulators $(S_{1}^{(M)},S_{2}^{(M)},s^{(M)})$ for each $M$, and
add every data point to each of the $(\log N)$ accumulators. One
therefore reduces the memory overhead to $\mathcal{O}(\log N)$ and
the runtime to $\mathcal{O}(N\log N)$~\citep{ambegaokar-ajp10}.

This solves the memory issue, but $\mathcal{O}(N\log N)$ can still
amount to a significant runtime overhead. We can however still improve
the runtime to $\mathcal{O}(N)$ by realizing that we can reuse the
bin results from the bin size $M$ for the bin size $2M$: instead
of adding each data point $x_{i}$ to every accumulator, we only add
it to the lowest level ($M=1$). Then whenever, we ``empty'' a bin
accumulator $s^{(M)}$ at the level $M$, we not only update the partial
sums $S_{p}^{(M)}$, but also add the result to the bin accumulator
at the next level: $s^{(2M)}\gets s^{(2M)}+s^{(M)}$~\footnote{During peer review, the author was made aware of unpublished work
by H.~G. Evertz, who implemented a similar strategy in 1990.}.

By the same argument as in Eq.~(\ref{eq:complex}), the logarithmic
binning analysis can thus be performed in $\mathcal{O}(N)$ time while
keeping the $\mathcal{O}(\log N)$ memory scaling. This is a negligible
overhead in most Monte Carlo calculations: even in the case of VAR(1)
model simulations (Sec.~\ref{sec:var1}), where the Monte Carlo updates
are next to trivial, the accumulation amounts to less than 10\% computing
time.

\section{Logarithmic Spectral Analysis\label{sec:spectrum}}

\subsection{Review: Spectrum estimates\label{sec:spectrum-review}}

As we have seen in the previous section, binning analysis allows us
to estimate the total effect of autocorrelation. However, ideally,
we would also like to estimate the most important individual contributing
time scales $\tau_{i}$ in Eq.~(\ref{eq:Cap}). In other words, we
would like to find an approximation:
\begin{equation}
c_{A}(m)\approx\sum_{i}C_{i}^{2}\alpha_{i}^{|m|}=\sum_{i}C_{i}^{2}\exp(-|m|/\tau_{i}),\label{eq:cAmapprox}
\end{equation}
where the index $i=1,...,S'$ runs over the dominant time scales.
(The tacit assumption $\alpha\ge0$ will be justified in Sec.~\ref{sec:mh}.)

Aside from analytical arguments, which are restricted to simple Markov
chains, one can use Eq.~(\ref{eq:Cest}) to get an estimate $c_{A}(p)$
for the autocorrelation function from and then attempt to fit a set
of exponential decays $\{(C_{i},\tau_{i})\}$ to it. An elegant way
of fitting is to employ the linear prediction method~\citep{prony,rahmn-ieee87}:
one first attempts to find an auxiliary $\mathrm{AR}(S')$ model (cf.~Sec.~\ref{sec:var1-tauint})
which best reproduces the observed $c_{A}(p)$, and then extracts
the $\{(C_{i},\tau_{i})\}$ using Eq.~(\ref{eq:cXvar1}). Once an
approximation (\ref{eq:cAmapprox}) is obtained, one can use Eq.~(\ref{eq:tauint})
to estimate the integrated autocorrelation time.

Since the method relies on the estimator (\ref{eq:Cest}), the drawbacks
outlined in Sec.~\ref{sec:binningrev} carry over to the spectrum
estimation: firstly, the unfavorable computational scaling of $\mathcal{O}(N)$
space and $\mathcal{O}(N\log N)$ time is also present here. Secondly,
linear prediction is susceptible to noise levels exceeding $\sim20\thinspace\mathrm{dB}$,
which means one again has to introduce a cutoff $W$ because of the
noise-to-signal problem (cf. Ref.~\onlinecite{deylon-pre17}).

\subsection{Spectral analysis from logarithmic binning}

Because of the computational expense of the conventional spectrum
estimators, we ideally would like estimate the spectrum from the logarithmic
binning analysis, which can be performed quickly, in $\mathcal{O}(N)$
time, and with little additional memory, $\mathcal{O}(\log N)$.

In order to do so, we first examine the influence of a single mode
with decay $\alpha$ on the binning analysis. For a single $c(m)=\alpha^{|m|}$,
Eq.~(\ref{eq:chiMcA}) gives:
\begin{equation}
T_{M}^{\prime}(\alpha):=\sum_{m}\Lambda_{M}(m)\,\alpha^{|m|}=\frac{1+\alpha}{1-\alpha}-\frac{2\alpha(1-\alpha^{M})}{M(1-\alpha)^{2}}.\label{eq:Tprime}
\end{equation}
The first term on the r.h.s. of Eq.~(\ref{eq:Tprime}) is the asymptotic
value, the integrated autocorrelation time, and we see that the difference
indeed drops as $\mathcal{O}(1/M)$, as observed in Sec.~\ref{sec:biascorr}.

\begin{figure}
\includegraphics[width=1\columnwidth]{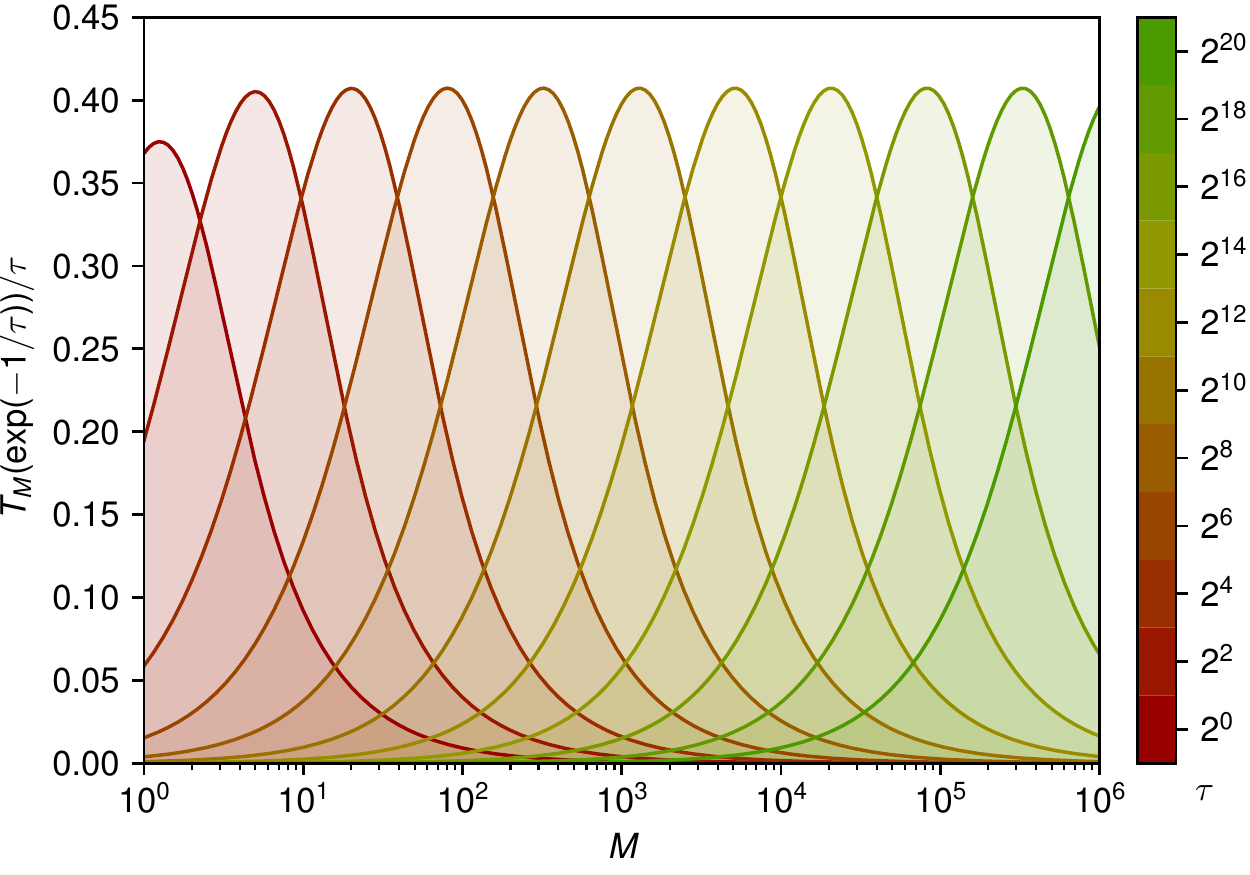}

\caption{Fitting function $T_{M}(\alpha)/\tau$ with $\alpha=\exp(-1/\tau)$
from Eq.~(\ref{eq:Tm}), plotted over $\log M$, where different
colors represent different values of $\tau$. The maximum of each
curve is at $M\approx1.28\tau$.}
\label{fig:tam}
\end{figure}

In order to analyze the spectrum, we remove the trivial term, and
define:
\begin{equation}
T_{M}(\alpha):=T_{2M}^{\prime}(\alpha)-T_{M}^{\prime}(\alpha)=\frac{\alpha(1-\alpha^{M})^{2}}{M(1-\alpha)^{2}}.\label{eq:Tm}
\end{equation}
Similarly, we combine the binning results as follows:
\begin{equation}
\theta_{M}[A]:=M(2\widehat{\Var}A^{(2M)}-\widehat{\Var}A^{(M)}).\label{eq:thetaM}
\end{equation}
Using Eqs.~(\ref{eq:Cap}) and (\ref{eq:chiMcA}), one then finds:
\begin{equation}
\theta_{M}[A]=\sum_{i}T_{M}(\alpha_{i})C_{i}^{2},\label{eq:thetatoci}
\end{equation}
with the coefficient $C_{i}=\langle Ab_{i}^{\prime}\rangle$.

The function $T_{M}(\exp(-1/\tau))$, normalized by the corresponding
$\tau$, is plotted in Fig.~\ref{fig:tam} over $\log M$. Different
curves correspond to different values of $\tau$. We see that each
value of $\tau$ can be translated into a Gaussian-like structure
in $M$ centered around $M_{\max}$, which is then observed by the
corresponding $\theta_{M}$. One can show that the maximum is located
at 
\begin{equation}
M_{\max}=[1+W(\ee^{-1})]\tau\approx1.28\tau,\label{eq:mmax}
\end{equation}
where $W(x)$ is the Lambert $W$ function.

Eq.~(\ref{eq:thetatoci}) is thus again a fitting problem, where
$\theta_{M}[A]$ is the observed data vector from logarithmic binning,
and $\alpha_{i}$ and $C_{i}$ are the model coefficients. Unlike
the case where we compute the spectrum from the autocorrelaiont estimator
(cf.~Sec.~\ref{sec:spectrum-review}), the data points are not equidistantly
spaced, and we unfortunately cannot use linear prediction or other
variants of Prony's method.

A simple but sufficiently accurate method to resolve the interdependency
in Eq.~(\ref{eq:thetatoci}) is to use a mesh for $\tau_{i}$ and
only fit the coefficients. Studying Fig.~\ref{fig:tam} and Eq.~(\ref{eq:mmax}),
we strive for ``equispaced'' curves in $\log M$. Therefore, we
make logarithmic ansatz for the decay times, i.e., 
\begin{equation}
\tau_{i}=r^{i},\qquad(r>1),\label{eq:tauansatz}
\end{equation}
and compute the corresponding $\alpha_{i}$ using Eq.~(\ref{eq:taui}).
Eq.~(\ref{eq:thetatoci}) can then be written as an ordinary least
squares problem:
\begin{equation}
\min_{x}||\theta-Tx||_{2},\label{eq:matvec}
\end{equation}
where $\theta_{i}$ is the vector of observations $\theta_{M_{i}}[A]$,
$T_{ij}$ is the fitting matrix $T_{M_{i}}(\alpha_{j})$, and $x_{j}$
is the vector of coefficients $C_{j}^{2}$.

On real binning data, it is important to use a generalized least squares
method~\citep{aitken-prsed34} instead of Eq.~(\ref{eq:matvec}).
This is because the statistical uncertainty in $\theta_{M}$ raises
as $\mathcal{O}(M)$ as evident from Fig.~\ref{fig:binning}b, so
the data points cannot be treated with equal weight. We empirically
found that one need not treat the full covariance matrix, and instead
approximate $\Sigma_{ij}\approx M_{i}\delta_{ij}$. The problem is
then reformulated as:
\begin{equation}
\min_{x}[(\theta-Tx)^{\mathrm{T}}\Sigma^{-1}(\theta-Tx)]=\min_{x}||\theta^{*}-T^{*}x||_{2},\label{eq:gls}
\end{equation}
where $\theta^{*}=\Sigma^{-1/2}\theta$ and $T^{*}=\Sigma^{-1/2}T$.

Both the ordinary (\ref{eq:matvec}) and the generalized least squares
problem (\ref{eq:gls}) are poorly conditioned. (The condition number
is usually of the order $10^{8}$.) To battle such a problem, regularization
techniques are commonly used, where a small regularization term $\lambda||x||_{p}$
is added to the cost function (\ref{eq:gls}). In our case, we found
that large values of $\lambda$ are needed to stabilize regressions
for $p\in\{1,2\}$, introducing significant systematic bias.

Instead, we exploit the fact that $C_{i}^{2}\ge0$ and use the non-negative
least squares method (NNLS), where we minimize Eq.~(\ref{eq:gls})
subject to a non-negativity constraint $x_{i}\ge0$~\citep{lawson-hanson-ls}.
We find that NNLS method is numerically stable when run on our fitting
problem. After the NNLS procedure, the fitted coefficients $x_{i}$
now give the approximation to the autocorrelation spectrum on a logarithmic
mesh $\tau_{i}$:
\begin{equation}
c_{A}(m)\approx\frac{1}{\Var A}\sum_{i}x_{i}\exp(-|m|/r^{i}).\label{eq:capprox}
\end{equation}

Once we have found an approximation (\ref{eq:capprox}) to the autocorrelation
function, we can use it to reconstruct a filtered version of the binning
analysis through Eq.~(\ref{eq:Tm}) and, more importantly, find an
estimate for the integrated autocorrelation time through Eq.~(\ref{eq:tauint})
that does not rely on visual inspection or an external parameter.

\subsection{Example: vector autoregression for $\alpha>0$\label{sec:var1}}

\begin{figure}
\includegraphics[width=1\columnwidth]{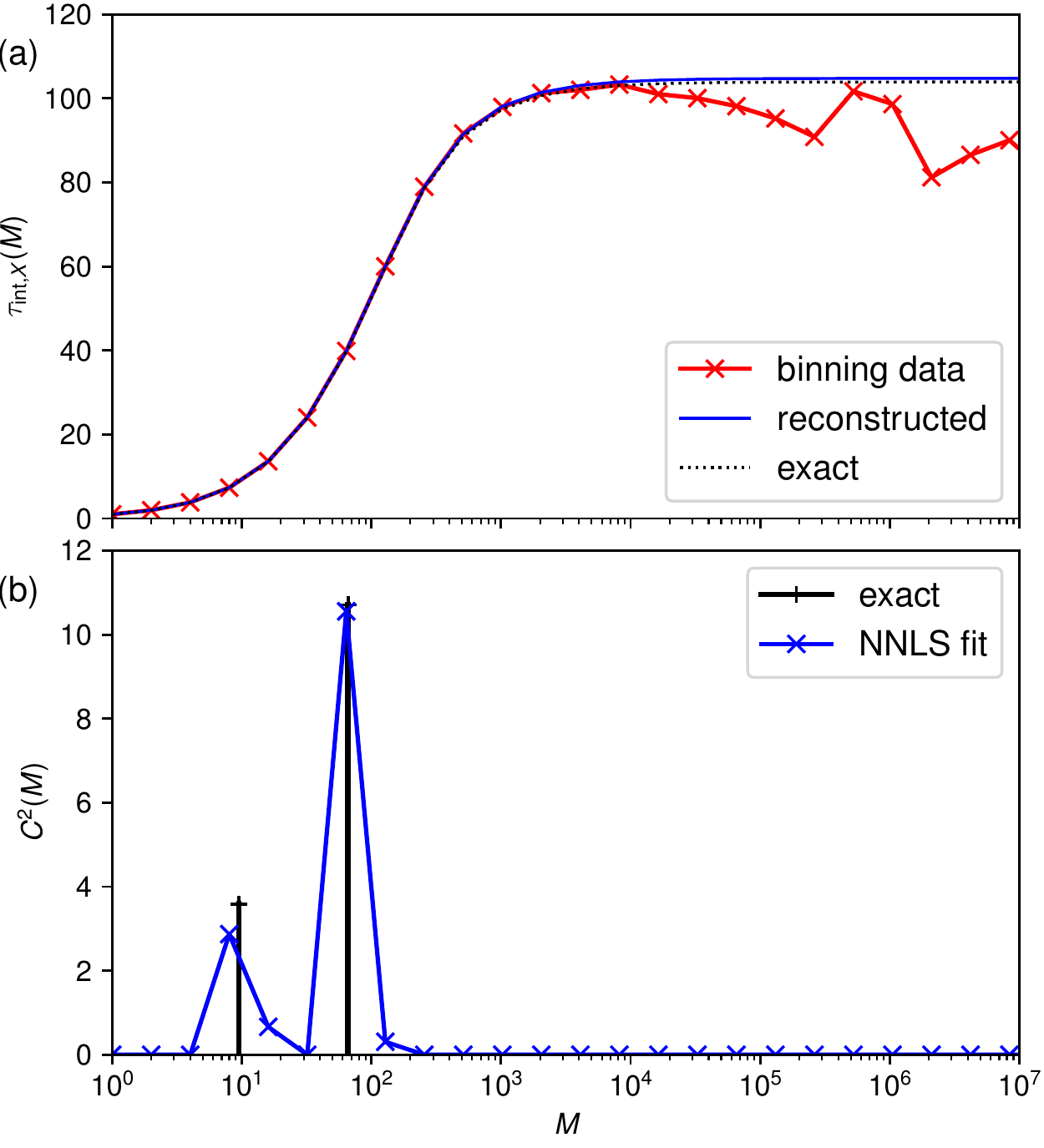}

\caption{Two-component VAR(1) model with eigenvalues $\alpha=(0.9,0.985)$,
where the first component $X_{1}$ is analyzed. (a) Logarithmic binning
analysis (cf.~Sec.~\ref{sec:binning}): binning data (red crosses)
from simulation of Eq. (\ref{eq:var1}) for $N=2^{26}$ steps, exact
behavior (black dotted curve) computed from Eq.~(\ref{eq:Tm}), and
curve reconstructed from spectral analysis (blue curve). (b) Spectral
analysis: coefficients $C_{i}^{2}$ for $\alpha_{i}=2^{i}$, estimated
from NNLS fit (blue crosses) and the exact peaks and weights from
Eqs.~(\ref{eq:cXvar1}) and (\ref{eq:sylv}) as black spikes.}
\label{fig:var1}
\end{figure}

As an example, we return to our VAR(1) model of Sec.~\ref{sec:var1-tauint}.
We again use two components ($n=2$) and generated the propagation
matrix $\phi_{1}$ from two eigenvalues $\alpha=(0.9,0.985)$, which
correspond to $\tau\approx(9.5,66.2)$. We rotated the eigenbasis
by $60$ degrees to introduce coupling between the components. From
Eq.~(\ref{eq:tauinty}), we find that the expansion coefficients
are $C_{i}^{2}\approx(3.59,10.71)$. We then generated $N=2^{26}$
configurations, and measured the first component $X_{1}^{(t)}$ for
each configuration.

The corresponding (biased) autocorrelation estimator from Eq.~(\ref{eq:tauest})
is shown in Fig.~\ref{fig:var1}a (red crosses). The exact curve
in dotted black is given for comparison. The corresponding spectral
analysis is given in Fig.~\ref{fig:var1}b. For simplicity, we chose
$r=2$ in Eq.~(\ref{eq:tauansatz}), which generates an identical
grid for binning analysis and spectral analysis. The NNLS fitted coefficients
$x_{i}$ are shown as blue crosses, and we can see that they closely
follow the analytic peaks in both location and magnitude.

The blue curve in Fig.~\ref{fig:var1}a shows the reconstructed binning
data, by plugging the model coefficients into Eq.~(\ref{eq:Tm}):
we see that the curve smoothens out the statistical uncertainties
at high $M$ due to the use of weights in the least squares procedure,
shows the correct monotonic behavior due to the non-negativity constraint
in the least squares procedure, and overall follows the analytic curve
(dotted black) closely. Similarly, computing the autocorrelation time
from the spectral analysis yields $\tauint X\approx104.38$ compared
to the true value of $\tauint X\approx103.90$.

\subsection{Metropolis--Hastings Monte Carlo\label{sec:mh}}

By making a logarithmic ansatz (\ref{eq:tauansatz}) for the autocorrelation
times and using non-negative least squares, we have tacitly assumed
that the Markov chain has no negative eigenvalues that contribute
to the estimator, i.e., $\alpha_{i}\ge0$. This is not true in general,
as can easily be seen by setting $\phi_{1}$ to a matrix with negative
eigenvalues in Eq.~(\ref{eq:var1}). Negative eigenvalues express
``cyclic'' behavior in the Markov chain, where we get alternating
correlation and anti-correlation as we move forward in Markov time.

Fortunately, one can show that the Markov chain for the Me\-tro\-po\-lis--Hastings
algorithm~\citep{metropolis-jcp-1953,hastings-biomet-1970} (at least
in its basic form) has no negative ei\-gen\-val\-ues~\citep{liu-statcmp96}.
This allows us to drop the absolute value in Eq.~(\ref{eq:taui})
and unambiguously interpret $\tau$ as time scale.

In simulations of Gibbs ensembles, these time scales are related to
barriers in phase space of height $\tau\propto\exp(\beta E)$, where
$\beta$ is the inverse temperature and $E$ is the energy difference
to the intermittent excited state. This means we can interpret $\log(\tau_{i})$
as the height of phase space barriers and the corresponding $x_{i}$
as the combined influence of barriers of that strength to the estimator
in question.

\section{Application: Ising model\label{sec:ising}}

In order to perform logarithmic spectral analysis on a realistic Metropolis--Hastings
Monte Carlo simulation, we consider the ferromagnetic Ising model.
The results for the autocorrelation spectra are well-known there,
allowing us to verify the results, while at the same time providing
a realistic testbed for the method.

The Hamiltonian of the Ising model is given by~\citep{LandauBinder05}:
\begin{equation}
\mathcal{H}=-\sum_{\langle ij\rangle}\sigma_{i}\sigma_{j},\label{eq:ising}
\end{equation}
where $\langle ij\rangle$ runs over all pairs of directly neighboring
Ising spins $\sigma_{i}\in\{1,-1\}$ on a $L\times L$ square lattice
with periodic boundary conditions. Since the system is finite and
there is no external magnetic field, $\langle m\rangle=0$ by symmetry.

We perform a Metropolis--Hastings Monte Carlo simulation for Eq.~(\ref{eq:ising})
with so called ``typewriter sweeps'': we go through the spins of
the lattice in rows from top to bottom, traversing each row from left
to right. For each spin we encounter, the proposed move is to flip
it $\sigma_{i}\to-\sigma_{i}$.

It is well known that with these kinds of updates, the Markov chain
gets ``stuck'' in one of two minima corresponding to a majority
spin-up or spin-down configuration. These minima are separated by
a phase space barrier of height $2\beta L$, which is the energy required
to form a grain boundary between spin-up and spin-down. Consequentially,
the results will show spurious polarization at low temperatures or
large system sizes.

\begin{figure}
\includegraphics[width=1\columnwidth]{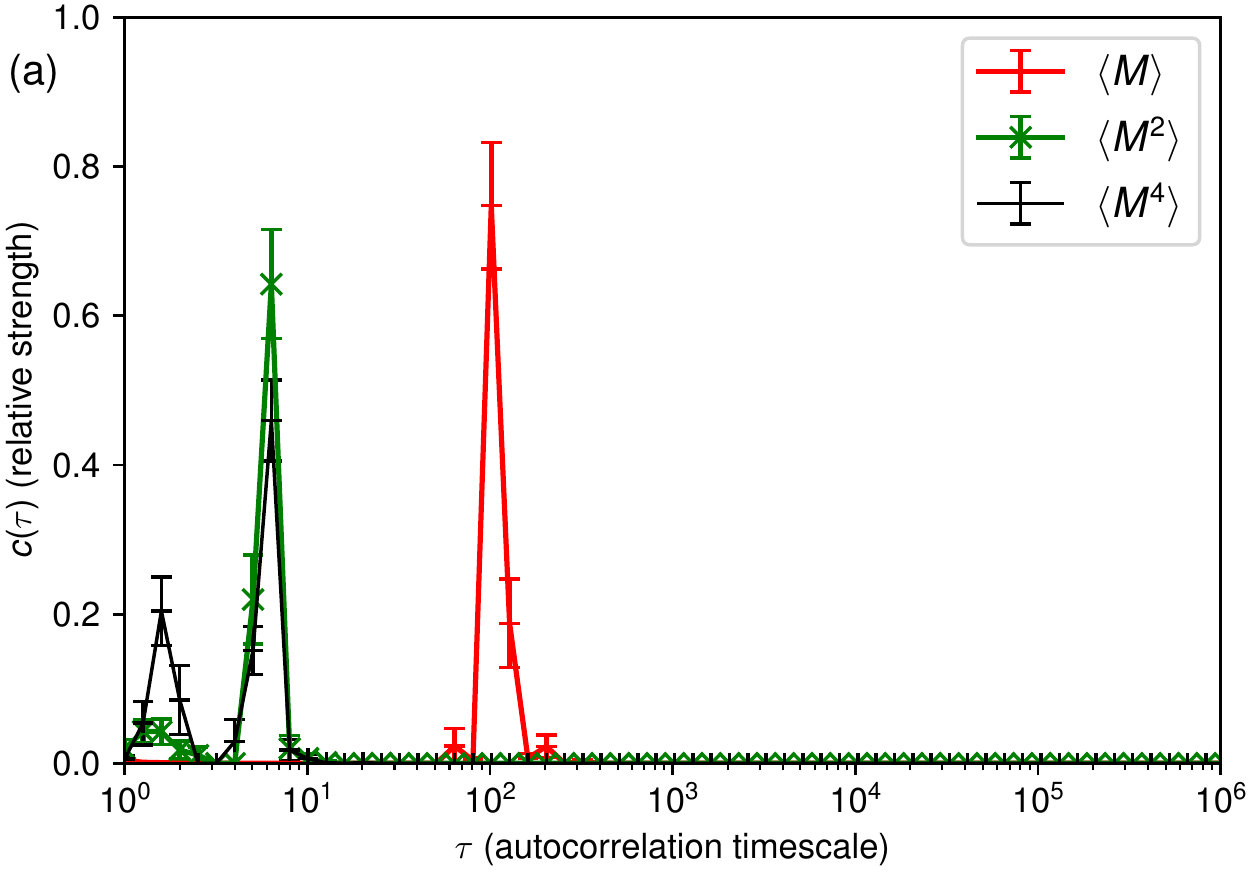}

\includegraphics[width=0.97\columnwidth]{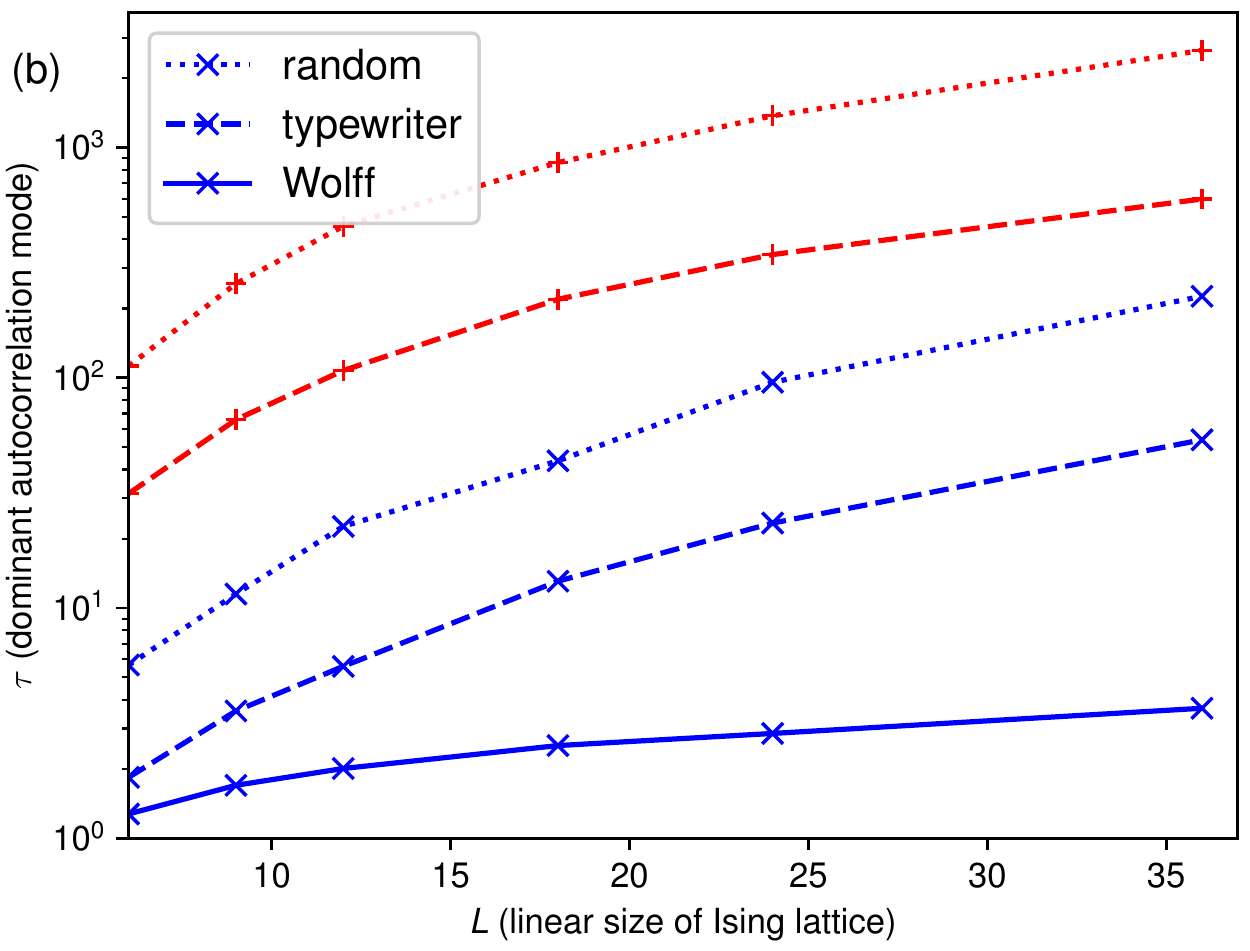}\caption{Ising model on a $L\times L$ square lattice for $\beta=1/2.3$ (a)
Spectral analysis for the mean magnetization $\langle M\rangle$ (red
plusses), $\langle M^{2}\rangle$ (green crosses), and $\langle M^{4}\rangle$
(black plusses), in the case $L=12$ for a Markov chain with typewriter
sweeps. (b) Scaling of the dominant autocorrelation mode $\tau$ with
linear system size $L$ for the magnetization (red plusses) and $\langle M^{2}\rangle$
(blue crosses) for three different kind of sweeps: set of $L^{2}$
random spin flips (dotted lines), typewriter sweeps (dashed lines),
and Wolff cluster updates (solid line).}
\label{fig:ising}
\end{figure}

These minima and their depth can be clearly seen on the autocorrelation
spectra in Fig.~\ref{fig:ising}a for the mean magnetization and
powers thereof, given by $\hat{M}^{\alpha}=(\sum_{i}\sigma_{i})^{\alpha}/L^{2}$.
We set the temperature to $T=2.3$, just above the critical temperature.
The curve for the mean magnetization $\langle M\rangle$ shows a peak
around $\tau\approx100$, which corresponds to significant phase space
barrier. This feature is absent in $\langle M^{2}\rangle$ and $\langle M^{4}\rangle$
identifying the source of the leading barrier as one that polarizes
the system.

This suggests the inclusion of a global move, which just flips all
spins and gets rid of the spurious polarization. To see whether such
a move would solve the autocorrelation problem, we plot the location
of the dominant autocorrelation mode over the linear system size $L$
for the typewriter updates (dashed curves in Fig.~\ref{fig:ising}b):
we see that there is a dominant autocorrelation mode in both $\langle M\rangle$
(red dashed curve) and $\langle M^{2}\rangle$ (blue dashed curve)
that scales similarly with system size. Thus, while a move that flips
all spins solves the immediate issue in $\langle M\rangle$, it does
not solve the autocorrelation problem in larger powers $\langle M^{2}\rangle$
and $\langle M^{4}\rangle$, since a global spin flip does not affect
$M^{2}$.

To make sure the problem is related to single spin flips, we compare
the scaling of the dominant autocorrelation mode for the ``typewriter
sweeps'' (dashed line), where we have a deterministic pattern of
proposing spins to flip, with ``random flip'' sweeps (dotted line),
where we choose $L^{2}$ spins at random with replacement and propose
to flip each one of them. It is well-known\citep{berg-mcmc04} that
typewriter sweeps outperform a set of random flips because of the
higher likelihood to propose clusters of spins, and this is evident
in an offset of the two curves in Fig~\ref{fig:ising}b. However,
we see that the scaling both for $\langle M\rangle$ (red dashed vs.
dotted curve) and $\langle M^{2}\rangle$ (blue dashed vs. dotted
curve) is similar, indicating a problem common to single spin flip
moves.

Eventually, the autocorrelation problem is solved by cluster updates~\citep{swensen-wang,Wolff89}.
Using the Wolff cluster updates~\citep{Wolff89}, we find that the
autocorrelation in $\langle M\rangle$ is trivially solved and the
dominant autocorrelation mode of $\langle M^{2}\rangle$ (blue solid
line) is considerably smaller in magnitude than when we use other
updates. More importantly, we see that the scaling of the autocorrelation
mode is more favorable, indicating we have overcome a phase space
barrier.

Let us emphasize again that our conclusions about the effectiveness
of the moves for Ising simulations are nothing ``new'': the Ising
model has a sufficiently simple structure and the Monte Carlo moves
have a simple physical interpretation, so that we can use our physical
intuition to analyze the structure of the Markov chain. The point
here is to illustrate that logarithmic spectral analysis provides
a way to obtain these properties without \emph{a priori }knowledge
or physical intuition at negligible additional computational cost.
It is thus suited for more complicated models or more advanced Monte
Carlo techniques.

\section{Conclusions}

In this paper we have shown how to perform logarithmic binning analysis
and spectral analysis to extract information about the autocorrelation
on Markov chain Monte Carlo runs. We have shown that these analyses
can be performed at modest additional computational cost and without
external parameters.

Logarithmic spectral analysis in particular allows quantitative and
statistically robust analysis of the types, height, and influence
of phase space barriers on the Monte Carlo result. It also can be
used to get improved estimates of the integrated autocorrelation time,
essential for postprocessing applications such as statistical hypothesis
testing~\citep{wallerberger-pre17}. It has the advantage of working
with the existing logarithmic binning analysis data already implemented
in many codes.

We feel confident that logarithmic spectral analysis will become an
integral part of developing or maintaining Markov chain Monte Carlo
codes: it provides a clear path on how to design and improve Markov
chain moves to avoid autocorrelation effects without relying on physical
intuition or trial-and-error alone, and comes at only modest additional
computational cost.

The techniques outlined here are straight-forward to implement, and
we encourage the reader to do so in order to get a deeper understanding.
For production codes, an optimized and tested implementation of these
techniques is scheduled for inclusion in the upcoming version of the
ALPS core libraries~\citep{gaenko-cpc17}.
\begin{acknowledgments}
The author would like to thank Jia Li, Igor Krivenko and Emanuel Gull
for fruitful discussions and careful review of the manuscript. MW
was supported by the Simons Foundation via the Simons Collaboration
on the Many-Electron Problem. This research used resources of the
National Energy Research Scientific Computing Center, a DOE Office
of Science User Facility supported by the Office of Science of the
U.S. Department of Energy under Contract No. DE-AC02-05CH11231.
\end{acknowledgments}

\end{document}